\documentclass[preprint,showpacs,amsfonts,amsmath,amssymb]{revtex4}
\usepackage{graphicx}
\usepackage{dcolumn}
\usepackage{bm}
\usepackage{showlabels}

\begin{document}
\title{ Geodesic Structure of Test Particle in Bardeen
Spacetime }
\author{Sheng Zhou}
\author{Juhua Chen} \email{jhchen@hunnu.edu.cn}
\author{Yongjiu Wang}
\affiliation{College of Physics and Information Science, Hunan
Normal University, Changsha, Hunan 410081, P. R. China}
\begin{abstract}
The Bardeen model describes a regular space-time, i.e. a singularity-free black hole space-time. In this paper, by analyzing the behavior of the
effective potential for the particles and photons, we investigate the time-like and null geodesic structures in the space-time of Bardeen model. At the same time, all kinds of orbits, which are allowed according to the energy level corresponding to the effective potentials,  are numerically simulated in detail. We find many-world bound orbits, two-world escape orbits and escape orbits in this spacetime. We also find that bound orbits precession directions are opposite and their precession velocities are different, the inner bound orbits shift along counter-clockwise with high velocity while the exterior bound orbits shift along  clockwise with low velocity.


\end{abstract}
\maketitle
\section{Introduction}
The Bardeen model describes a regular space-time  \cite{Bardeen,Borde} using the energy-momentum tensor of nonlinear electrodynamics as the source of the
field equations and it is also known as a regular black-hole solution which obeys the weak energy condition. This global regularity of black hole solutions is quite important to understand the final state of gravitational collapse of initially regular configurations. When  ratio of mass to
charge is $27g^2\leqslant16m^2$, the Bardeen model represents
a black hole and a singularity-free structure \cite{Eloy1}. When
$27g^2=16m^2$, the horizons shrink into a single one, which corresponds to an extreme black hole
such as  the extreme Reissner-Nordstr\"{o}m solution. The physically reasonable source for regular black hole solution to
Einstein equations has been reported around 1998 \cite{Eloy2,Eloy3,Eloy4,Magli}. In the Bardeen model,
the parameter $g$ representing the magnetic charge of the nonlinear self-gravitating
monopole\cite{Eloy1}, was studied later on.

It is well known that many effects, such as bending of light, gravitational time-delay, gravitational red-shift and precession of  planetary orbits,  were predicted by General Relativity. Because these gravitational effects are very important for theories and observations, many theoretical physics and astrophysics are interested in investigating them for different gravitational systems.  The  geodesic structure with a positive cosmological constant was investigated by Jaklitsch et al.\cite{Jaklitsch}, the corresponding effective potential was analyzed in detail.
The analysis of the effective potential for null geodesics in the Reissner-Nordstr\"{o}m-de Sitter
and Kerr-de Sitter space-time was carried out in Refs. \cite{Stuchlik} and \cite{Jiao}.
All possible geodesic motions in the extreme Schwarzschild-de Sitter
space-time were investigated by Podolsky \cite{Podolsky}. Lake investigated light deflection in the
Schwarzschild-de Sitter space-time\cite{Lake}. Exact solutions in closed analytic form for the
geodesic motion in the Kottler space-time were considered by Kraniotis et al \cite{Kraniotis1}. Kraniotis \cite{Kraniotis2} investigated
the geodesic motion of a massive particle in the Kerr and Kerr (anti)de Sitter gravitational field by solving
the Hamilton acobi partial differential equation.  Cruz et al.\cite{Cruz}
studied the geodesic structure of the Schwarzschild anti-de Sitter black hole. Chen
and Wang \cite{chen1,chen2,chen3,chen4} have investigated the orbital dynamics
of a test particle in  gravitational fields with an electric dipole and a mass quadrupole, and
in the extreme Reissner-Nordstr\"{o}m black hole spacetime. The motion of test particle in Ho$\breve{r}$ava-Lifshitz black hole space-times
was studied using numerical techniques \cite{Enolskii}.

To find all of the possible orbits which are allowed by the energy levels for time-like and null geodesic in Bardeen spacetime, we   analysis the effective potentials in detail. To  describe the trajectories of massive and null particles, we have a direct visualization of
the allowed motions. This paper is organized as follows: In Section II, we give a brief review on the Bardeen spacetime.
In Section III, we give out the motion equations, and define the effective potential. In Section IV and V, we discuss the time-like and null geodesic structure of the Bardeen spacetime in detail. A conclusion is given in the last section.

\section{The Bardeen spacetime}
The line element representing the Bardeen spacetime is given by\cite{Borde}
\begin{eqnarray}\label{metric}
 ds^2=&-&[1-\frac{2mr^2}{(r^2+g^2)^{\frac{3}{2}}}]dt^2+[1-\frac{2mr^2}{(r^2+g^2)^{\frac{3}{2}}}]^{-1}dr^2 \nonumber \\&+&r^2(d\theta^2+sin^2\theta d\phi^2),
\end{eqnarray}
where the parameter g represents the magnetic charge
of the nonlinear self-gravitating monopole\cite{Eloy1}. The corresponding lapse function is
\begin{eqnarray}
 f(r)=1-\frac{2mr^2}{(r^2+g^2)^{\frac{3}{2}}}.
\end{eqnarray}
The Bardeen model describes a regular space-time for the following inequality:
\begin{eqnarray}
 g^2\leqslant\frac{16}{27}m^2.
\end{eqnarray}
When $g^2<\frac{16}{27}m^2$, there are two horizons in Bardeen spacetime.
For the equality $g^2=\frac{16}{27}m^2$, the horizons shrink into a single one, which are showed in Fig.1 in detail.

\begin{figure}[h!]
\begin{center}
\includegraphics[scale=0.6]{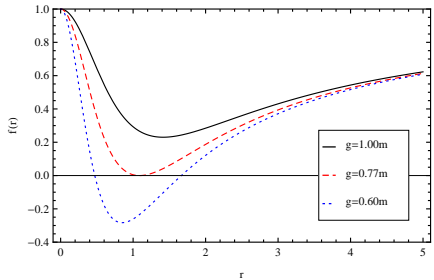}
\end{center}
\caption{Horizons of the  Bardeen spacetime.}
  \end{figure}

\section{Geodesics equation}
It is well known that the Euler-Lagrange equations for the variational
problem associated to spacetime  metric  describes the geodesics. So we set up the corresponding Lagrangian according to Eq.(\ref{metric})
\begin{eqnarray}\label{eq:Lagrangian}
\mathcal{L}=-f(r)\dot{t}^2+f(r)^{-1}\dot{r}^2+r^2(\dot{\theta}^2+sin^2\theta \dot{\phi}^2),
\end{eqnarray}
in which the dots denote the derivative with respect to the affine
parameter $\tau$. The Hamiltonian motion equations  are
\begin{equation}
\dot{\Pi}_q-\frac{\partial{\mathcal {L}}}{\partial q}=0 ,
\end{equation}
where $\Pi_q=\partial {\mathcal {L}}/\partial \dot{q}$ is the momentum to coordinate $q$. Since the Lagrangian is independent of $(t,\phi)$, the corresponding conjugate momentums are conserved, therefore
\begin{equation}\label{eq:E}
\Pi_t = -(1-\frac{2mr^2}{(r^2+g^2)^{\frac{3}{2}}})\dot{t} = -E,
\end{equation}
\begin{equation}\label{eq:L}
\Pi_\phi = r^2sin^2\theta\dot{\phi} = L,
\end{equation}
where $E$ and $L$ are motion constants.

From the motion equation  for $\theta$
\begin{eqnarray}
\dot{\Pi}_\theta-\frac{\partial{\L}}{\partial \theta}=0,
\end{eqnarray}
we obtain
\begin{equation}
\frac{d(r^2\dot{\theta})}{d\tau} = r^2 sin\theta cos\theta \dot{\phi}^2.
\end{equation}

If we simplify the above equation by choosing the initial conditions $\theta = \pi/2$, $\dot{\theta} = 0$ and $\ddot{\theta}= 0$,
the Eq.(\ref{eq:L}) becomes
\begin{equation}\label{eq:L1}
\Pi_\phi = r^2\dot{\phi} = L,
\end{equation}
from Eqs.(\ref{eq:E}, \ref{eq:L}), the Lagrangian (\ref{eq:Lagrangian}) can be written in the following form
\begin{equation}
2{\mathcal {L}}\equiv h = \frac{E^2}{1-\frac{2mr^2}{(r^2+g^2)^{\frac{3}{2}}}}-
\frac{\dot{r}^2}{1-\frac{2mr^2}{(r^2+g^2)^{\frac{3}{2}}}}-\frac{L^2}{r^2}.
\end{equation}
Now we solve the above equation for $\dot{r}^2$ in order to obtain the radial equation, which allows us to characterize possible moments of test particles and explicit solutions of the motion equation of test particles in the invariant plane
\begin{equation}\label{eq:motion}
\dot{r}^2 = E^2 - (1-\frac{2mr^2}{(r^2+g^2)^{\frac{3}{2}}})(h+\frac{L^2}{r^2}),
\end{equation}

It is useful to rewrite the above motion equation as a one-dimensional problem
\begin{equation}
\dot{r}^2 = E^2 - V_{eff}^2,
\end{equation}
where $V_{eff}^2$ is defined as an effective potential
\begin{equation}\label{eq:effective potential}
V_{eff}^2 = (1-\frac{2mr^2}{(r^2+g^2)^{\frac{3}{2}}})(h+\frac{L^2}{r^2}).
\end{equation}

\section{Time-like Geodesic Structure}
For time-like geodesic $h=1$, the corresponding effective potential becomes
\begin{equation}
V_{eff}^2 = (1-\frac{2mr^2}{(r^2+g^2)^{\frac{3}{2}}})(1+\frac{L^2}{r^2}).
\end{equation}
\begin{figure}[h!]
\begin{center}
\includegraphics[scale=0.6]{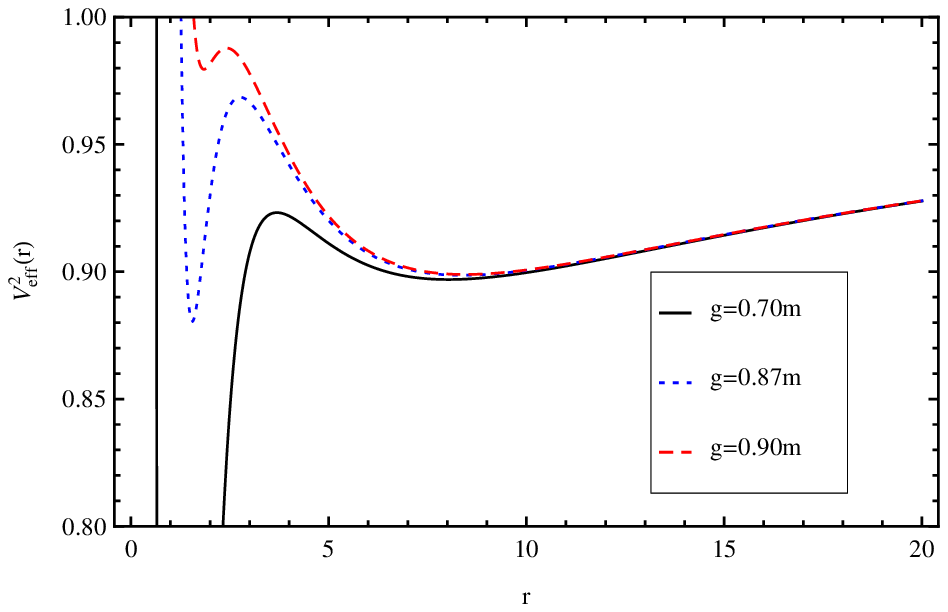}
\includegraphics[scale=0.6]{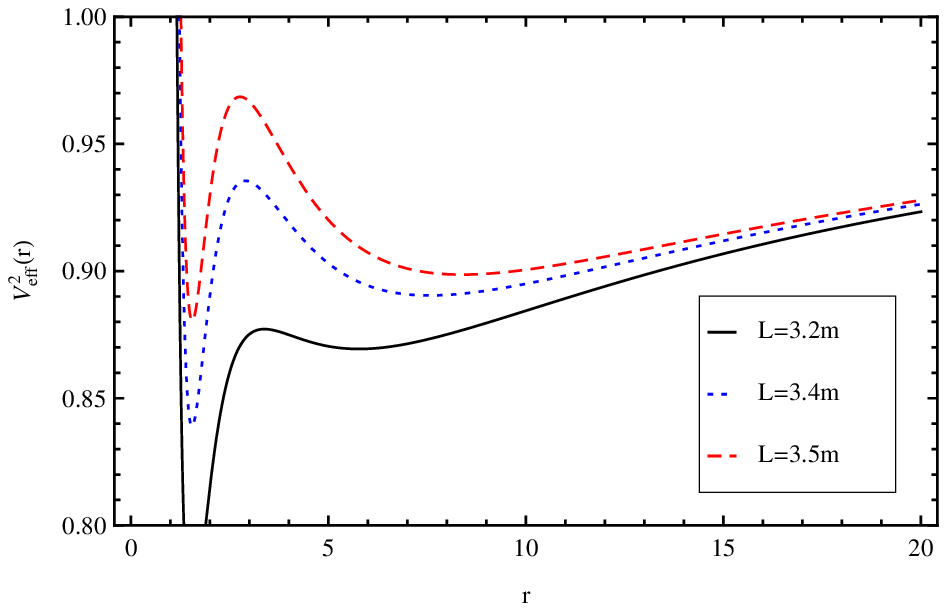}
\caption{The behavior of the effective potential of non-radial particle for fixed $L=3.5m$(left) and fixed $g=0.87m$(right).}
\end{center}
  \end{figure}
and the orbit equation for massive particle is
\begin{equation}\label{eq:time-like motion}
\dot{r}^2 = E^2 - (1-\frac{2mr^2}{(r^2+g^2)^{\frac{3}{2}}})(1+\frac{L^2}{r^2}).
\end{equation}

By using Eq.(\ref{eq:L1}) and making the
change of variable $u^{-1}=r$, we can obtain orbit equation for
massive particle
\begin{equation}\label{eq:time-like orbit1}
(\frac{du}{d\phi})^{2}= \frac{E^2-1}{L^2}-u^2+\frac{2mu+2mu^3L^2}{L^2(1+u^2g^2)^\frac{3}{2}},
\end{equation}
Differentiating (\ref{eq:time-like orbit1}), we have its second order motion equation
\begin{equation}\label{eq:time-like orbit2}
\frac{d^2u}{d\phi^2}+u=\frac{3mu^2L^2+3m}{L^2(1+u^2g^2)^\frac{5}{2}}-\frac{2m}{L^2(1+u^2g^2)^\frac{3}{2}},
\end{equation}

We solved (\ref{eq:time-like orbit1}) and (\ref{eq:time-like orbit2})
numerically to find all types of geodesics and examine how the
parameters influence on the timelike geodesics in the space-time of Bardeen model in detail.

From the effective potential curve (see Fig.3), we can identify 3 classes of orbits: i.e. planetary orbits, escape orbits and circular orbits when the energy of particle $E$ satisfies two critical values $E_{C_1}$ and $E_{C_2}$.
\begin{figure}[h!]
\begin{center}
\includegraphics[scale=0.7]{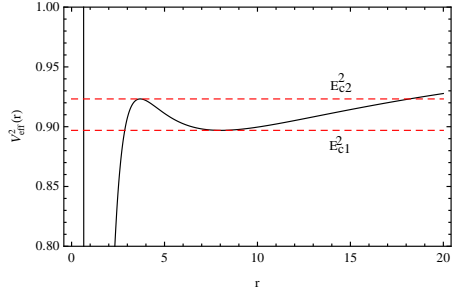}
\end{center}
\caption{The behavior of the effective potential in Bardeen space-time with $g=0.70m$, $L=3.5m$, $m=1$ and energy levels $E_{C_2}^2=0.90$ and $E_{C_1}^2=0.92$.}
  \end{figure}
\subsection{Time-like bound geodesics}
\begin{figure}[h!]
\begin{center}
\includegraphics[scale=0.7]{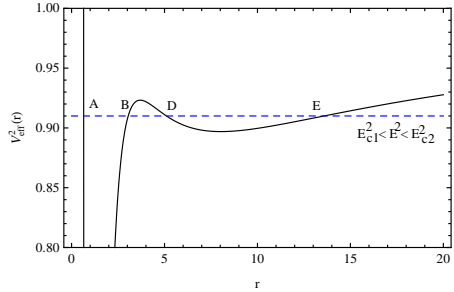}
\end{center}
\caption{The behavior of the effective potential of time-like bound geodesics for $E^2=0.91, g=0.7m, L=3.5m, m=1$}
  \end{figure}

In Fig.4 the dashed line denotes the value of the energy $E^2=0.91$, i.e. $E_{C_1}^2<E^2<E_{C_2}^2$. From potential curve, we can find two kinds of bound orbits for this energy level:

I) The particle orbits on a many-world bound orbit between the range $r_{A}<r<r_{B}$, which
is near the singularity and  can cross the two event horizons.  The $r_{A}$ and $r_{B}$
are the perihelion and aphelion distance of the planetary orbits, respectively. We also can find the clockwise precession of  planetary orbits which it is a well-known gravitational effect in general relativity theory.

II) The particle orbit is on a two-world bound orbit in the range $r_{D}<r<r_{E}$, where
the $r_{D}$ and $r_{E}$ are the perihelion and aphelion distance, which are
larger than the orbit of Case I. The two-world bound orbit is outside the event horizon. However, we also can find that the precession direction of  the planetary orbit  is counter-clockwise, and the precession velocity is slower than the orbit of Case I.

These two kinds of two-world bound orbits are simulated in Fig.6.

\begin{figure*}[h!]
\begin{center}
\includegraphics[scale=0.40]{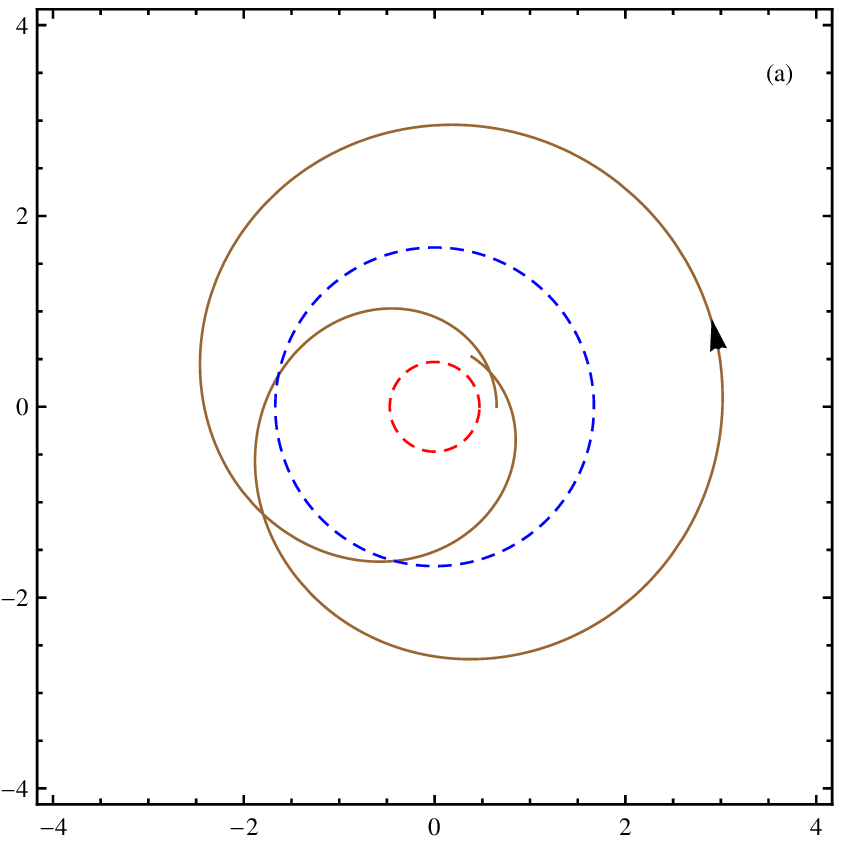}
\includegraphics[scale=0.40]{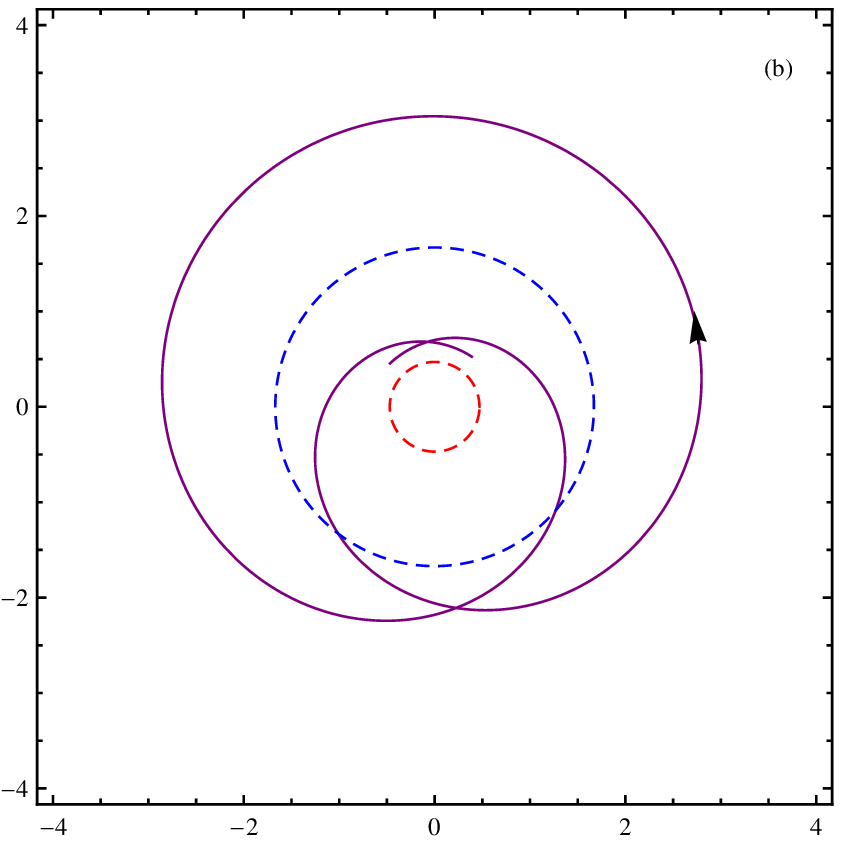}
\includegraphics[scale=0.40]{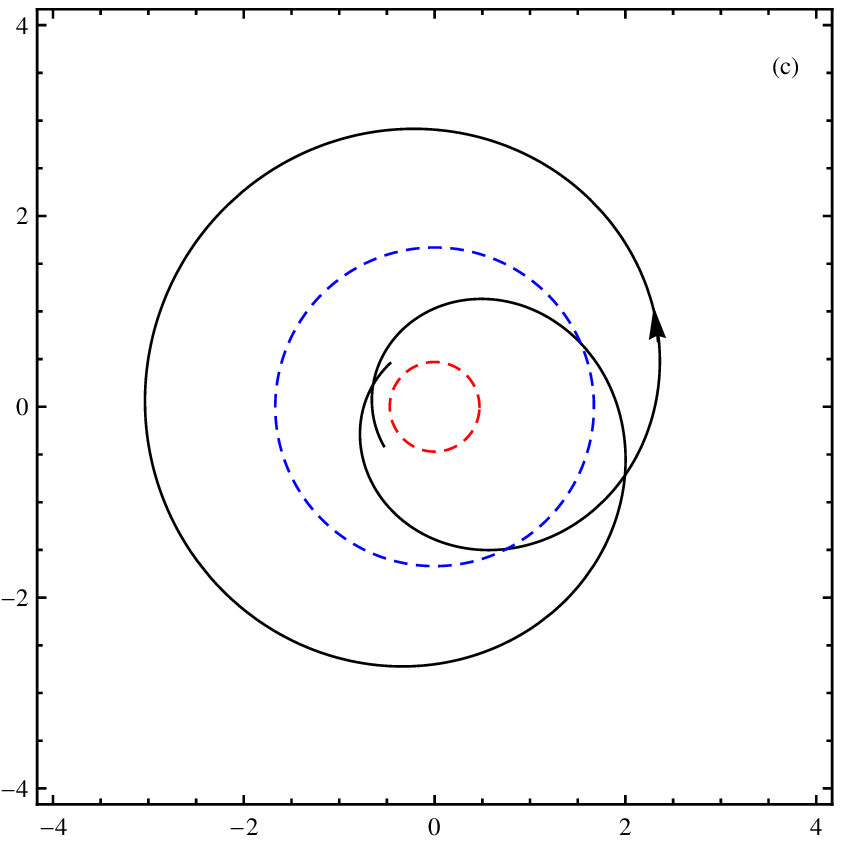}
\includegraphics[scale=0.40]{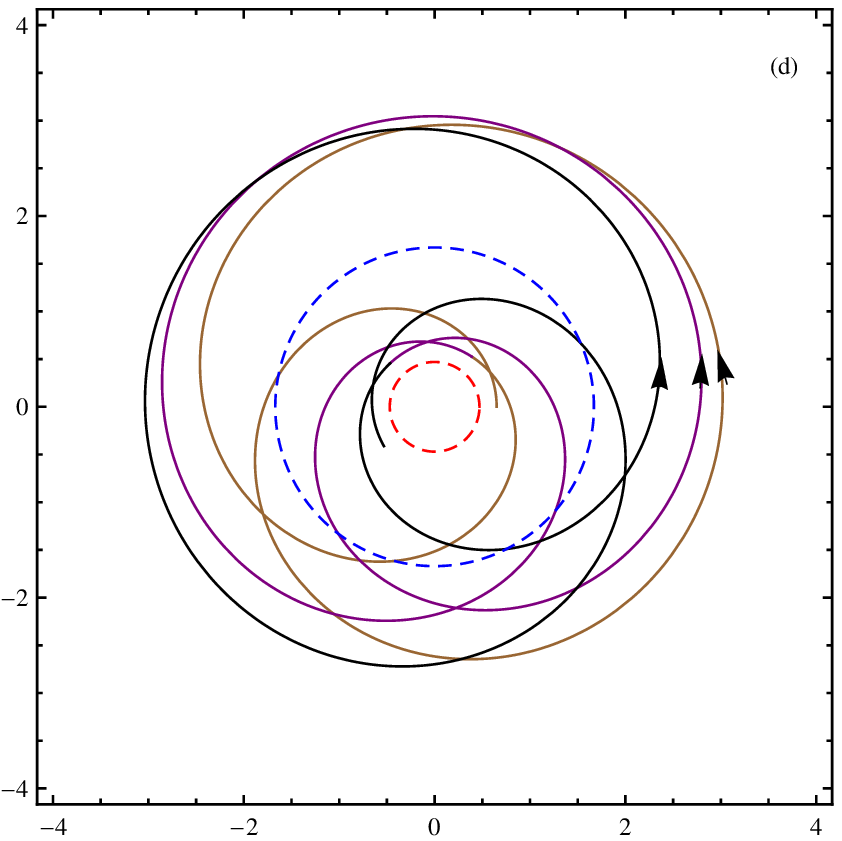}
\includegraphics[scale=0.40]{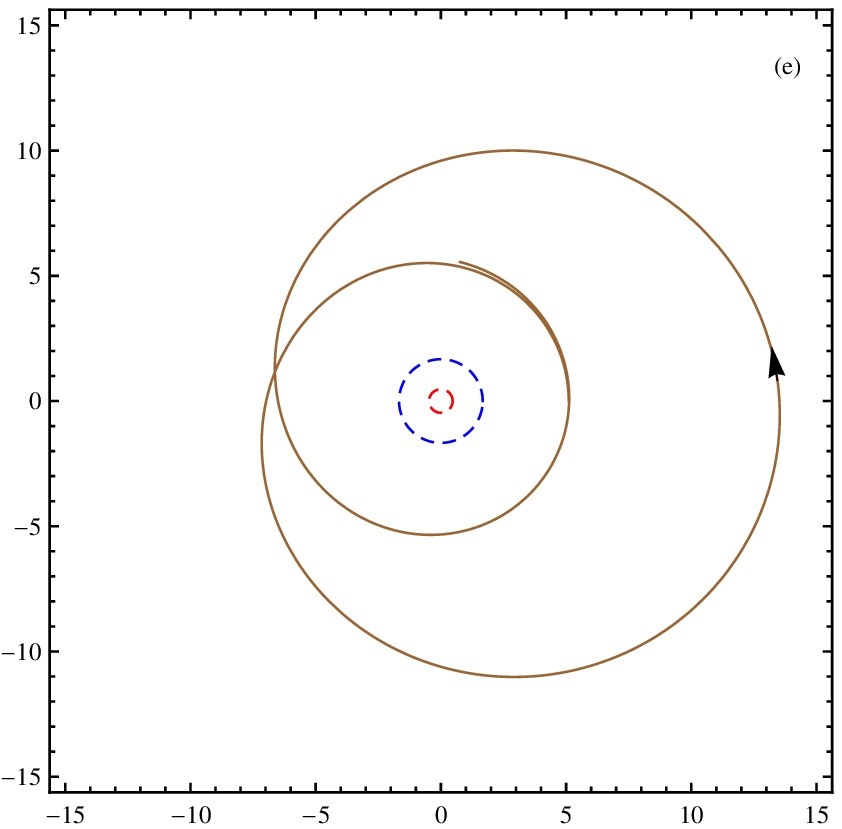}
\includegraphics[scale=0.40]{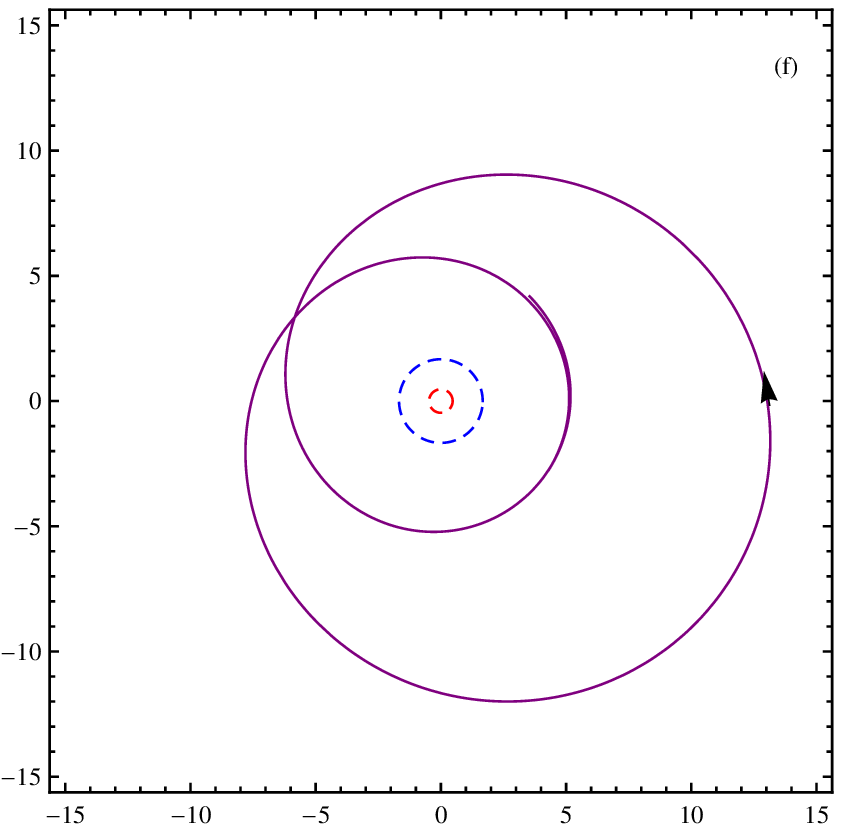}
\includegraphics[scale=0.40]{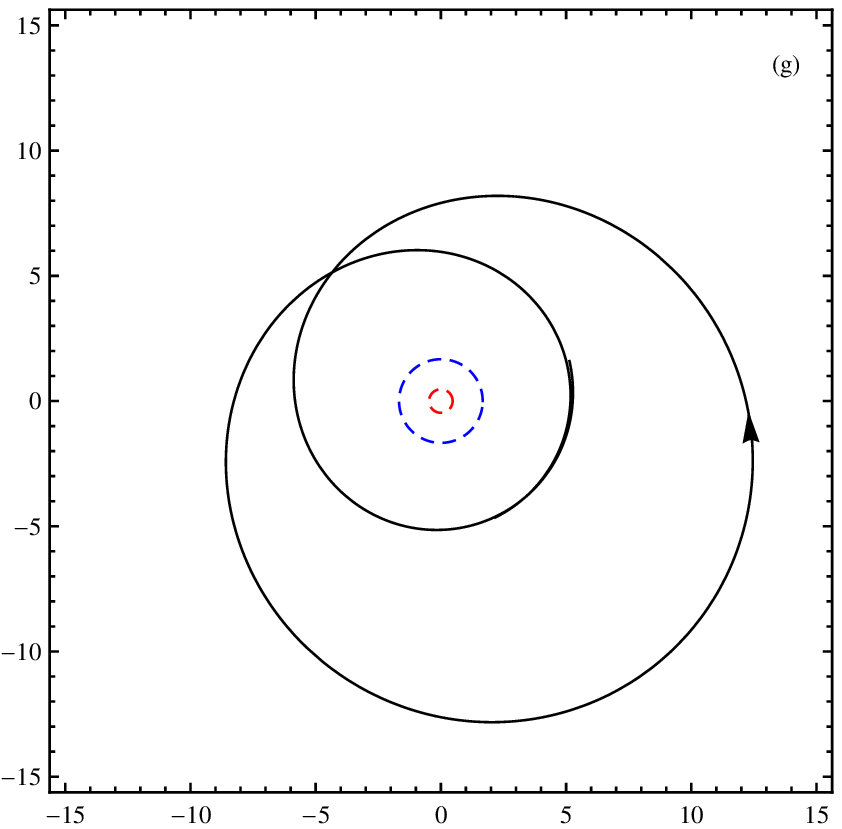}
\includegraphics[scale=0.40]{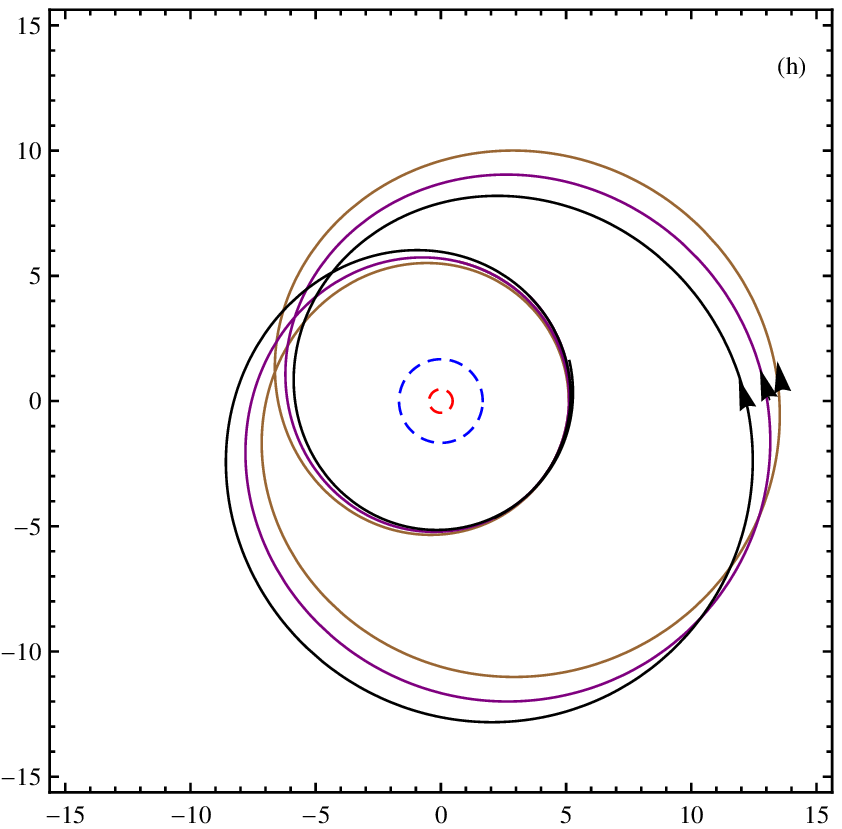}
\end{center}
\caption{Examples of the two-world bound orbit in the Bardeen space-time with $E^2=0.91$, $g=0.7m$, $L=3.5m$ and $m=1$.}
  \end{figure*}

\subsection{Time-like circle geodesics}
From Fig.6 and 7, we can see that there are two different circular  orbits. One is a unstable
circular orbit, the other one is a stable circular orbit.

I) When the energy of particle $E$ equals the peak value $E_{C_2}$ of the  effective potential, the particle can orbit on a unstable
circular orbit at $r=r_{C_2}$. Any perturbation would make such unstable orbit recede from $r=r_{C_2}$ to  $r=r_A$, then reflect at $r=r_A$, or move from $r=r_{C_2}$ to  $r=r_B$, then reflect at $r=r_B$. The particle will move between  $r=r_A$ and $r=r_B$ and  will make a unstable choice on a movement direction due to the perturbation. Figure 6 shows two cases numerically.

II) When the energy of particle $E$ equals the bottom value $E_{C_1}$ of the  effective potential, the particle can orbit on a stable
circular orbit at $r=r_{C_1}$. Or the particle orbits on a many-world bound orbit in the range $r_{A}<r<r_{D}$, where
the $r_{A}$ and $r_{D}$ are the perihelion and aphelion distance, respectively. Figure 7 shows two cases numerically.

\begin{figure*}[h!]
\begin{center}
\includegraphics[scale=0.60]{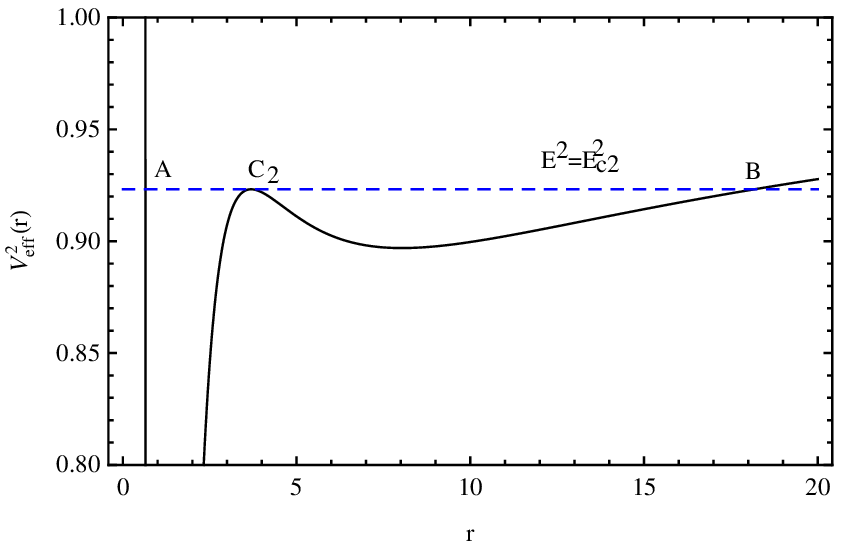}
\includegraphics[scale=0.40]{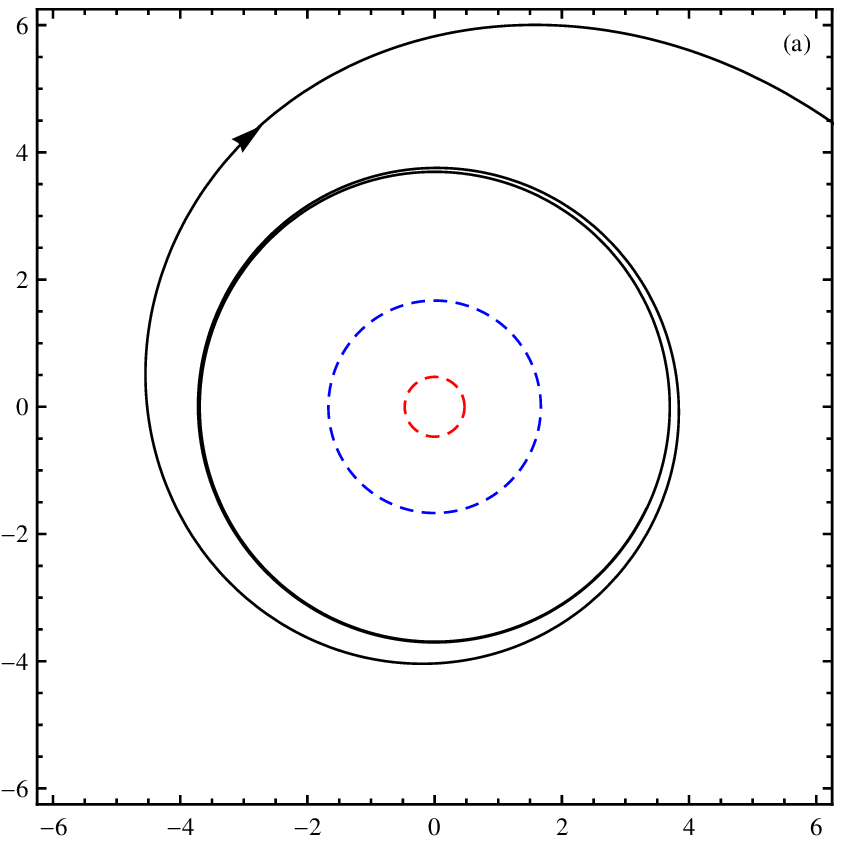}
\includegraphics[scale=0.40]{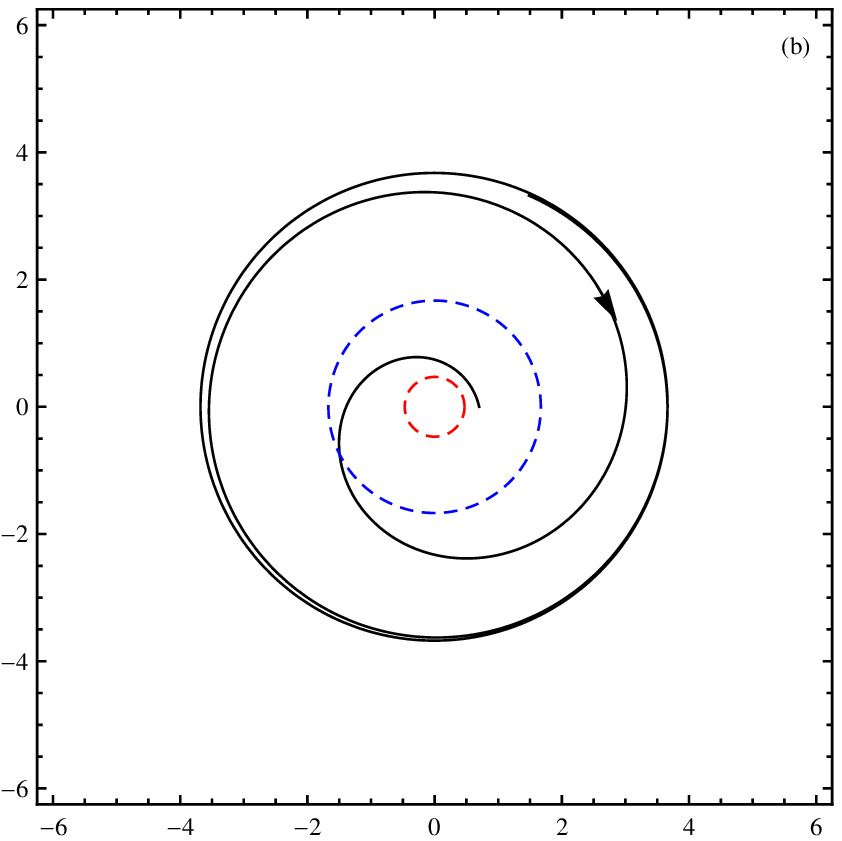}
\caption{Examples of the unstable time-like  circle orbit in the Bardeen space-time with $E_{C_2}^2=0.92$, $g=0.7m$, $L=3.5m$ and $m=1$.}
\end{center}
\begin{center}
\includegraphics[scale=0.60]{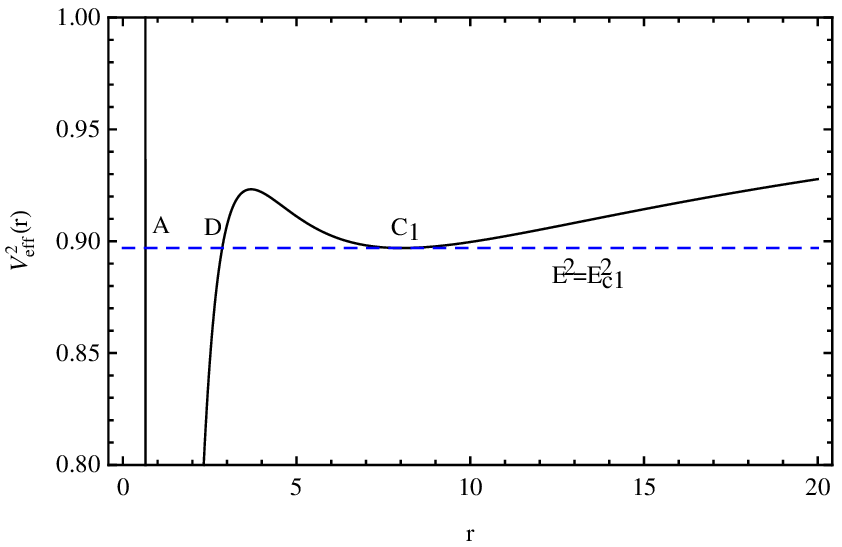}
\includegraphics[scale=0.40]{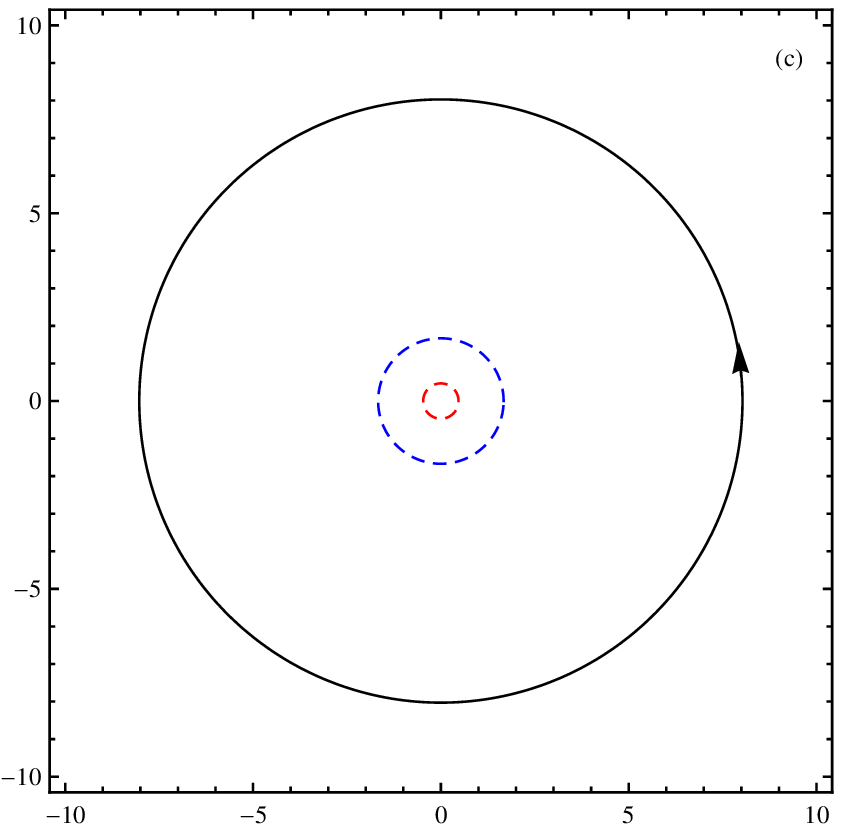}
\includegraphics[scale=0.40]{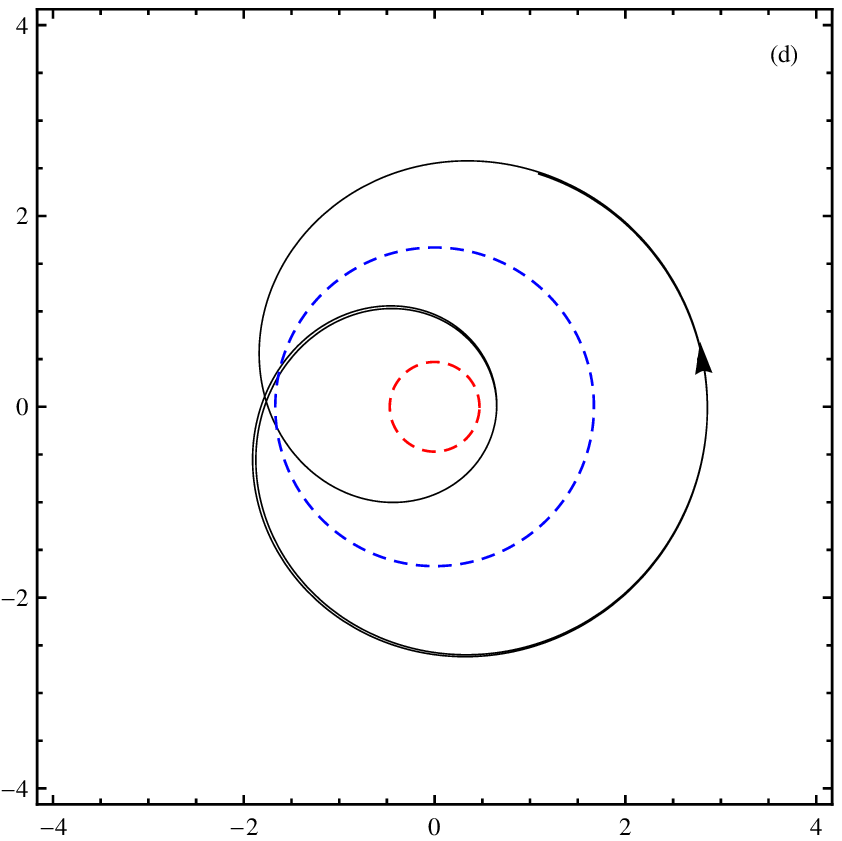}
\caption{Examples of the stable time-like circle orbit in the Bardeen space-time with $E^2=0.90$, $g=0.7m$, $L=3.5m$ and $m=1$.}
\end{center}
\end{figure*}

\subsection{Time-like escape geodesics}
When the particle energy is above the critical value (i.e. the peak value $E_{C_2}$ of the  effective potential), the particle can orbit on a two-world escape orbit with a curly structure and cross the two horizons  which is showed in Fig.8a. When the energy of particle  is much higher than the critical value, the escape orbit straightly deflects without curls, which is showed in Fig.8b. This means that the test particle coming from infinite would be reflected at a value of $r$ and would not be able to reach $r=0$, due to the infinite potential barrier at $r=0$. on the other words the particle approaches the black hole from an asymptotically flat region, crosses the horizons twice and moves away into another asymptotically region.

\begin{figure*}[h!]
\begin{center}
\includegraphics[scale=0.6]{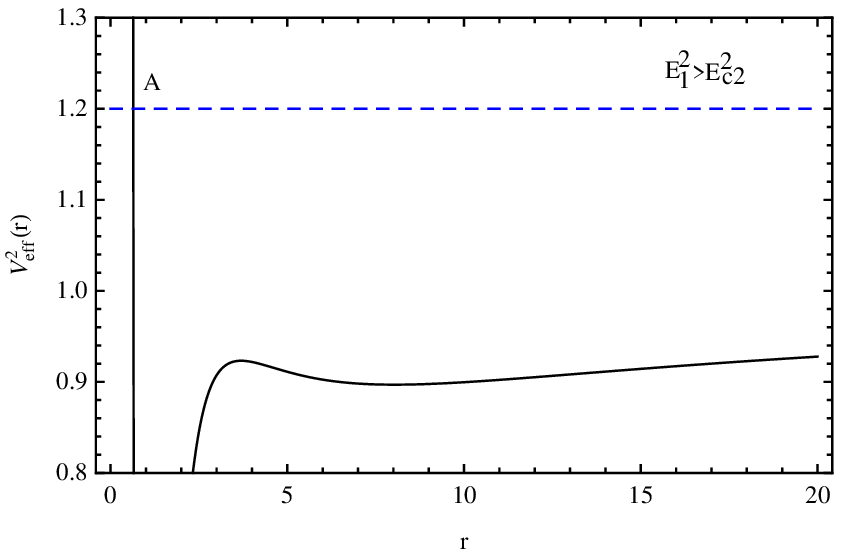}
\includegraphics[scale=0.40]{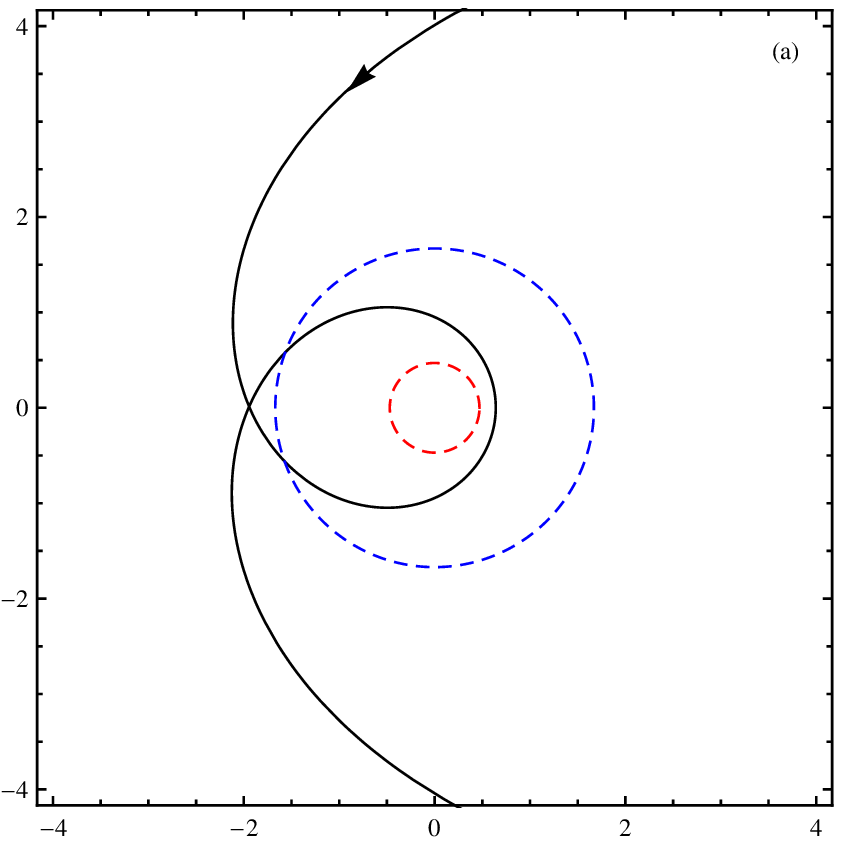}\\
\includegraphics[scale=0.6]{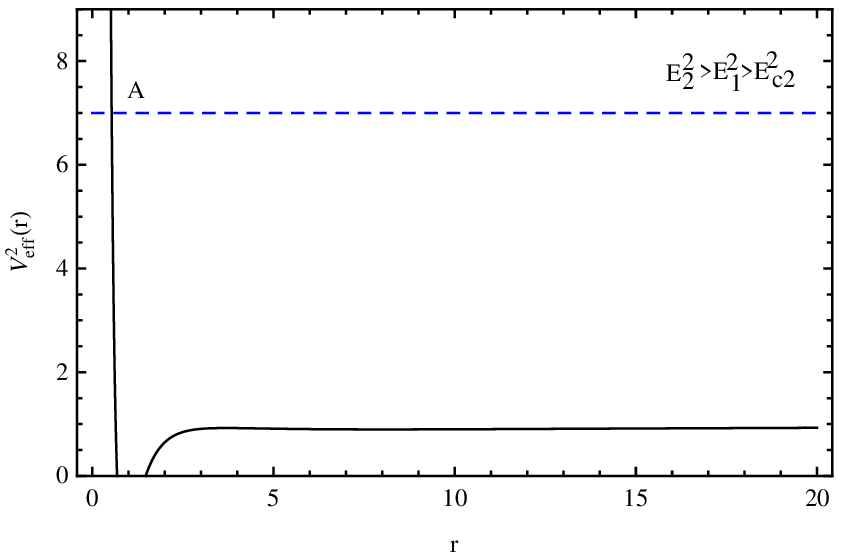}
\includegraphics[scale=0.40]{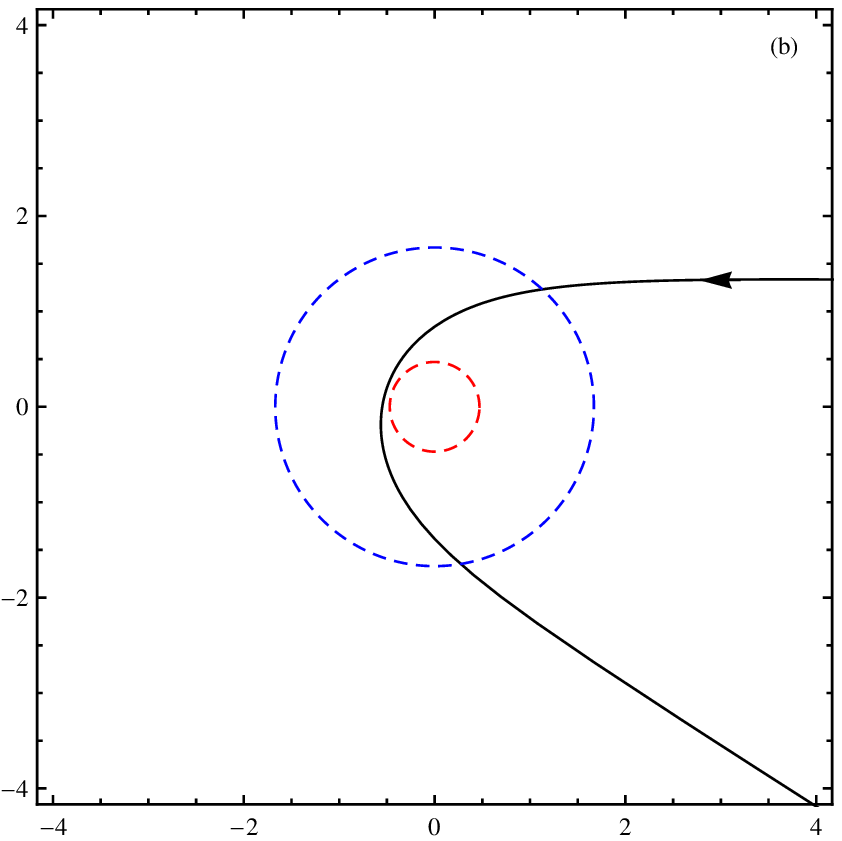}
\caption{Examples of the two-world escape orbit in the Bardeen space-time with $E_1^2=1.2$, $g=0.7m$, $L=3.5m$ and $m=1$ (top) and the time-like escape orbit in the Bardeen space-time with $E_2^2=7$, $g=0.7m$, $L=3.5m$ and $m=1$(bottom).}
\end{center}
\end{figure*}

\section{Null Geodesics}
For the null geodesic $h=0$, we get the corresponding effective potential from Eq.(\ref{eq:effective potential})
\begin{equation}
V_{eff}^2 = (1-\frac{2mr^2}{(r^2+g^2)^{\frac{3}{2}}})\frac{L^2}{r^2}.
\end{equation}
\begin{figure*}[htbp]
\begin{center}
\includegraphics[scale=0.60]{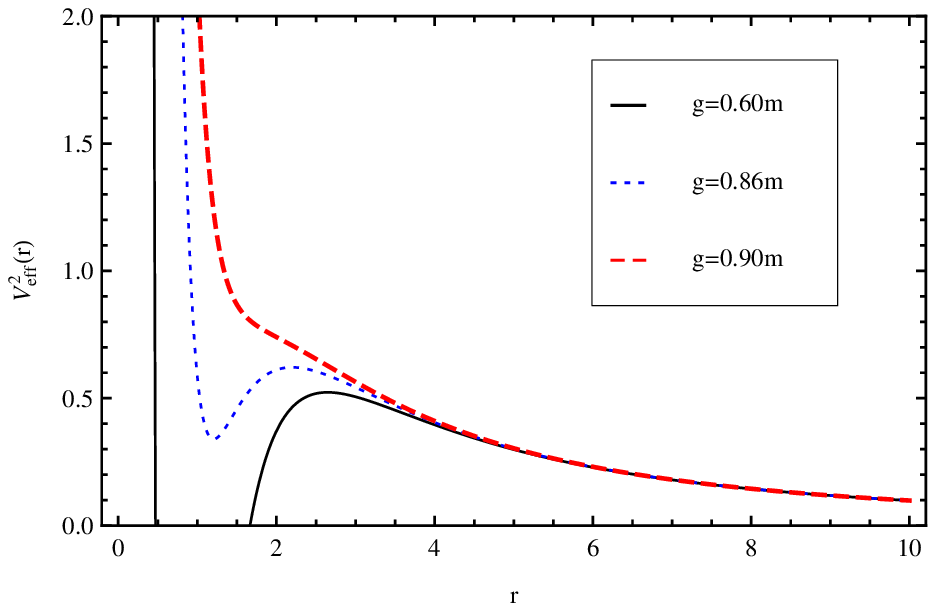}
\includegraphics[scale=0.60]{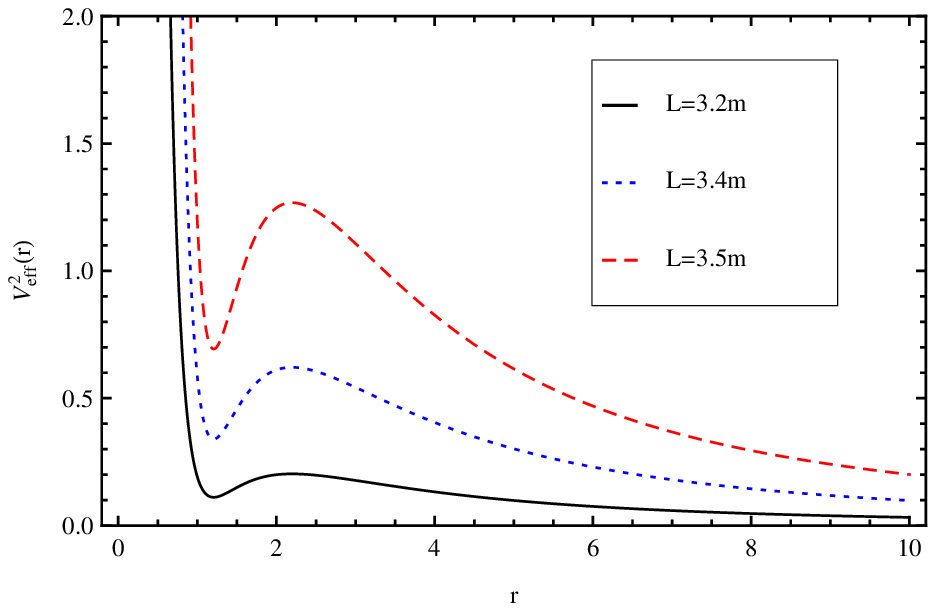}
\end{center}
\caption{The behavior of the effective potential of the null geodesics for fixed $L=3.5m, m=1(left)$ and for fixed $g=0.8m, m=1(right)$.}
\end{figure*}
The behavior of the effective potential depends on the parameters $g$, $L$ and the corresponding orbit equation  is
\begin{equation}\label{eq:null motion}
\dot{r}^2 = E^2 - (1-\frac{2mr^2}{(r^2+g^2)^{\frac{3}{2}}})\frac{L^2}{r^2}.
\end{equation}
By using Eq.(\ref{eq:null motion}) and making the change of variable $u^{-1}=r$, we obtain the orbit equation for
massive particle
\begin{equation}\label{eq:null orbit1}
(\frac{du}{d\phi})^{2}= \frac{E^2}{L^2}-u^2(1-\frac{2mu}{(1+u^2g^2)^\frac{3}{2}}).
\end{equation}
By differentiating the Eq.(\ref{eq:null orbit1}), we have
\begin{equation}\label{eq:null orbit2}
\frac{d^2u}{d\phi^2}+u=\frac{3mu^2}{(1+u^2g^2)^\frac{5}{2}}.
\end{equation}

We must solve the geodesic equations (\ref{eq:null orbit1}) and (\ref{eq:null orbit2})
numerically to investigate the null geodesics structure and how the
space-time parameters influence on the null geodesics structures in the Bardeen spacetime. We continue to follow the similar process of the Section IV.

\subsection{Null bound geodesics}
From the effective potential curve for photons in Fig.10, we can see that there are two different types of orbit when the energy $E$ belows the peak energy value $E_C$. When the initial position is between $r_A$ and $r_B$, the particle will move on a many-world bound orbit with the range of radius  from $r_A$ to
$r_B$. When the particle initial position is on the right hand side of the potential barrier, the particle approaches $r_D$ from an asymptotically flat region, then will be reflected to move away into another asymptotically region. These two kinds of orbits corresponding to the energy level are plotted on the right side of Fig.10, respectively.

\begin{figure*}[h!]
\begin{center}
\includegraphics[scale=0.60]{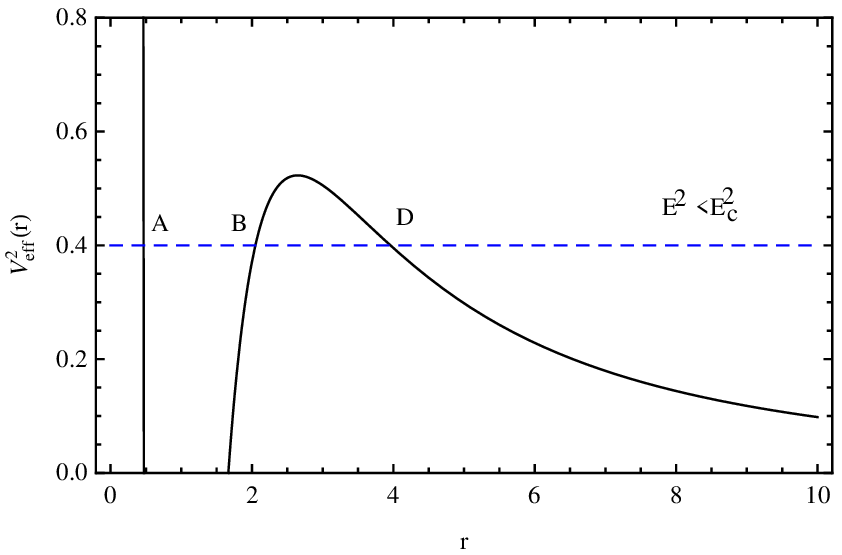}
\includegraphics[scale=0.40]{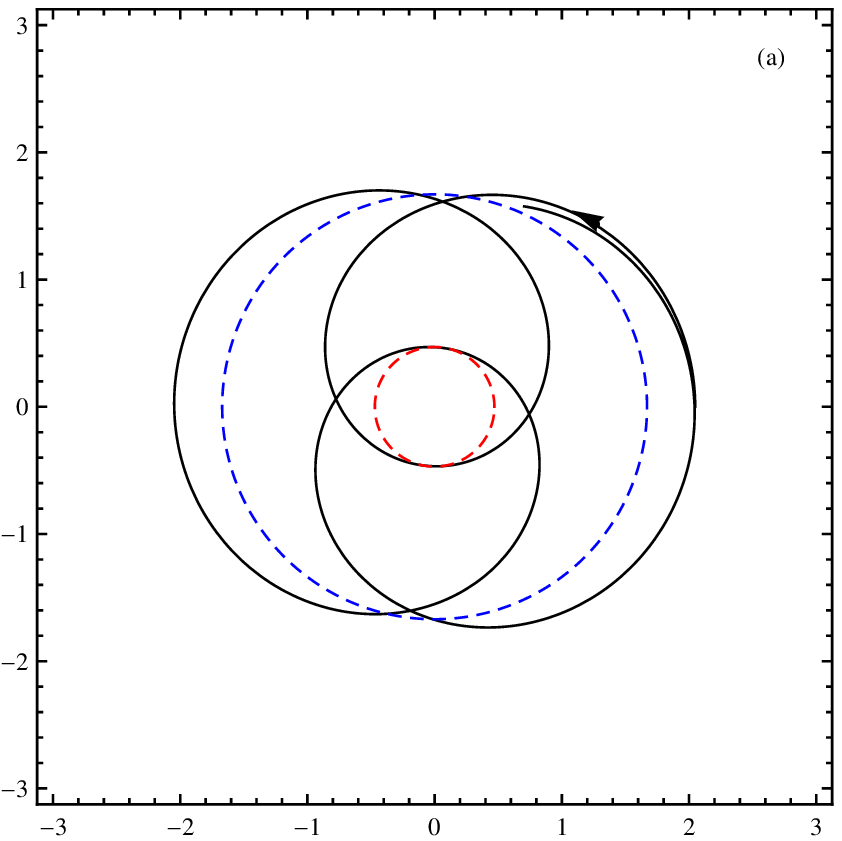}
\includegraphics[scale=0.40]{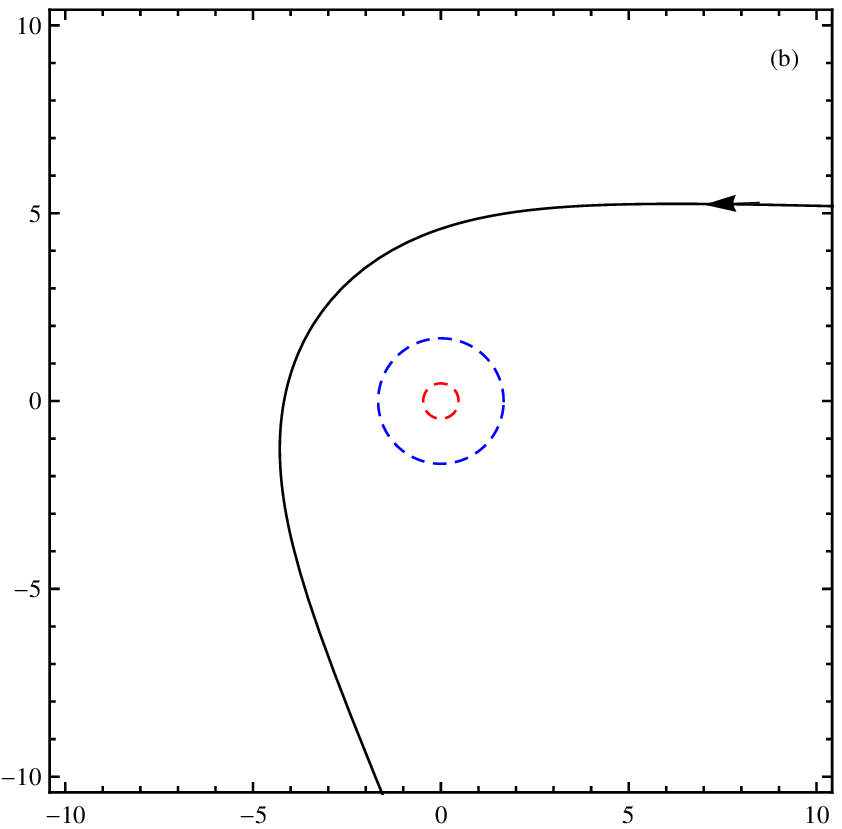}
\caption{Examples of of the many-world null bound and escape geodesics in  Bardeen spacetime with  $E^2=0.4$, $g=0.6m$, $L=3.5m$ and $m=1$.}
\end{center}
\end{figure*}

\subsection{Null circle geodesics}
When the energy $E=E_C$, The photon can orbit on a unstable circular orbit at $r=r_C$, and the photo on such
orbit will more likely recede from $r_C$ to $r_A$ crossing the horizons
 and will be reflected at  $r_A$ or escape to the infinity on the other side of the
potential barrier due the initial conditions and outside perturbation. Examples of such two kinds of orbits are shown in Fig.11.

\begin{figure*}[h!]
\begin{center}
\includegraphics[scale=0.60]{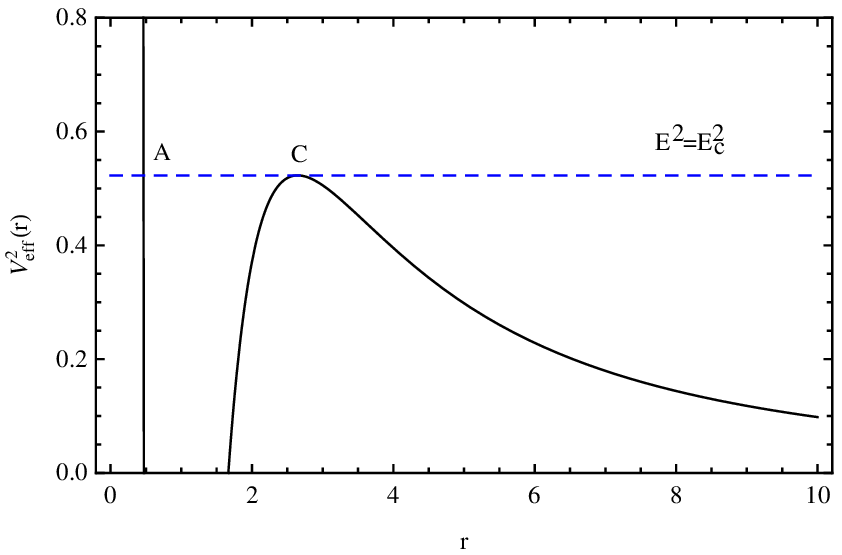}
\includegraphics[scale=0.40]{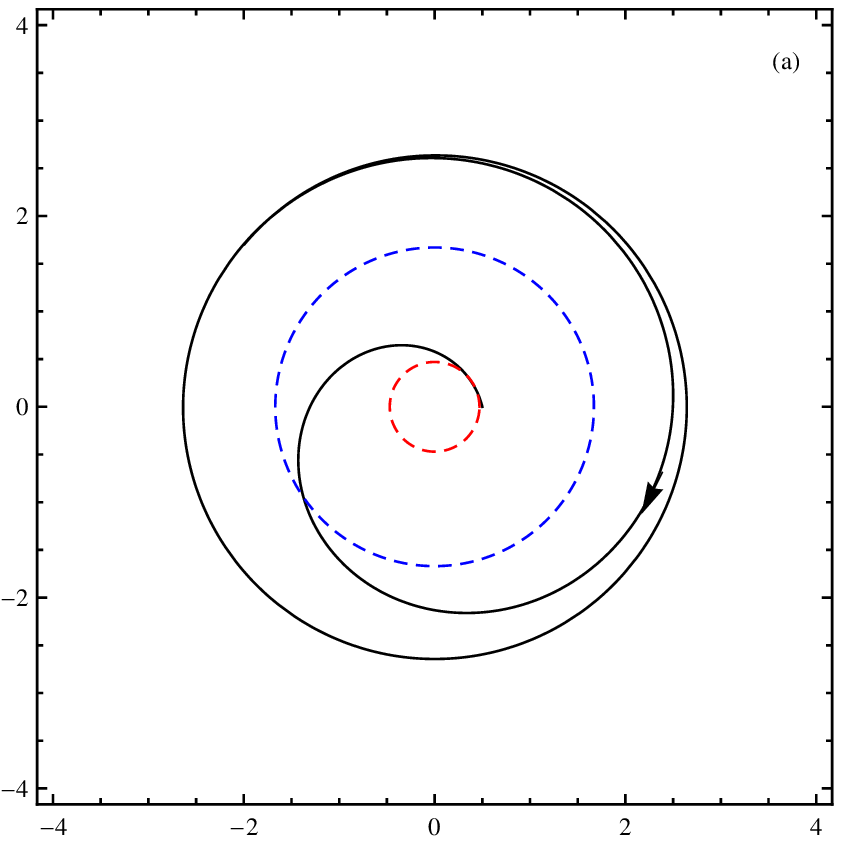}
\includegraphics[scale=0.40]{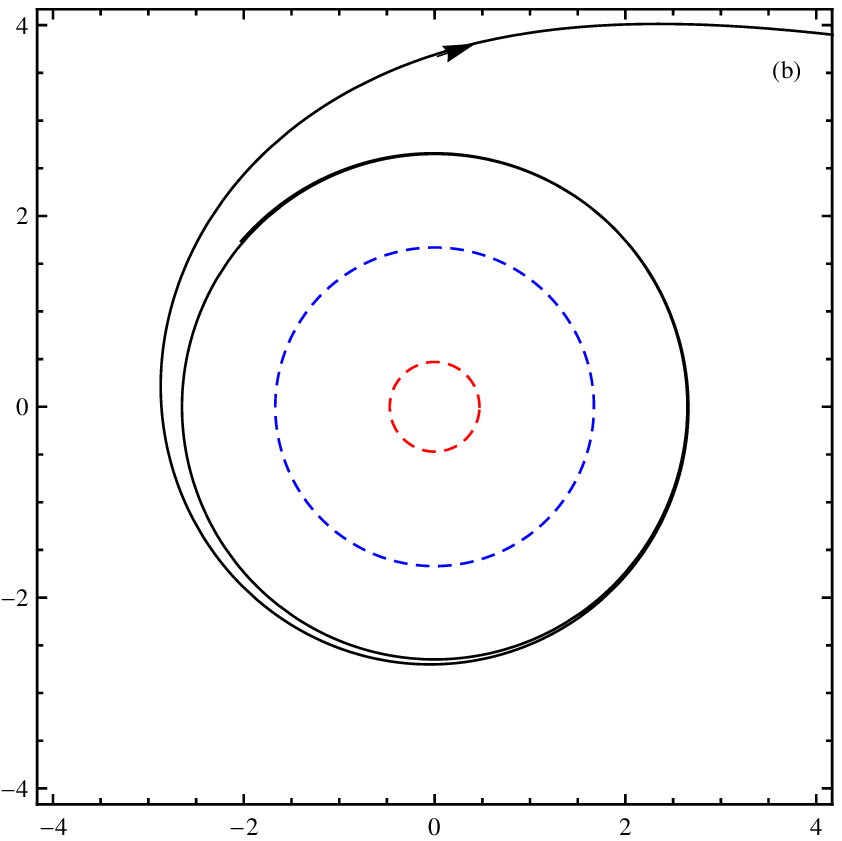}
\caption{Examples of of two kinds of unstable null circle geodesics in  Bardeen spacetime with  $E^2=0.52$, $g=0.6m$, $L=3.5m$ and $m=1$.}
    \end{center}
\end{figure*}

\subsection{Null escape geodesics}
When the energy $E>E_C$, The photon will be on the three different kinds of the escape geodesics which are shown in Fig.12. We can see that when the
energy level becomes larger from 0.6 to 7, the corresponding orbit changes from the two-world escape orbit to  the escape orbit without intersection point. this means the particle approaches the black hole from an asymptotically flat region, crosses the horizons and then moves away into another asymptotically region.
\begin{figure*}[h!]
\begin{center}
\includegraphics[scale=0.45]{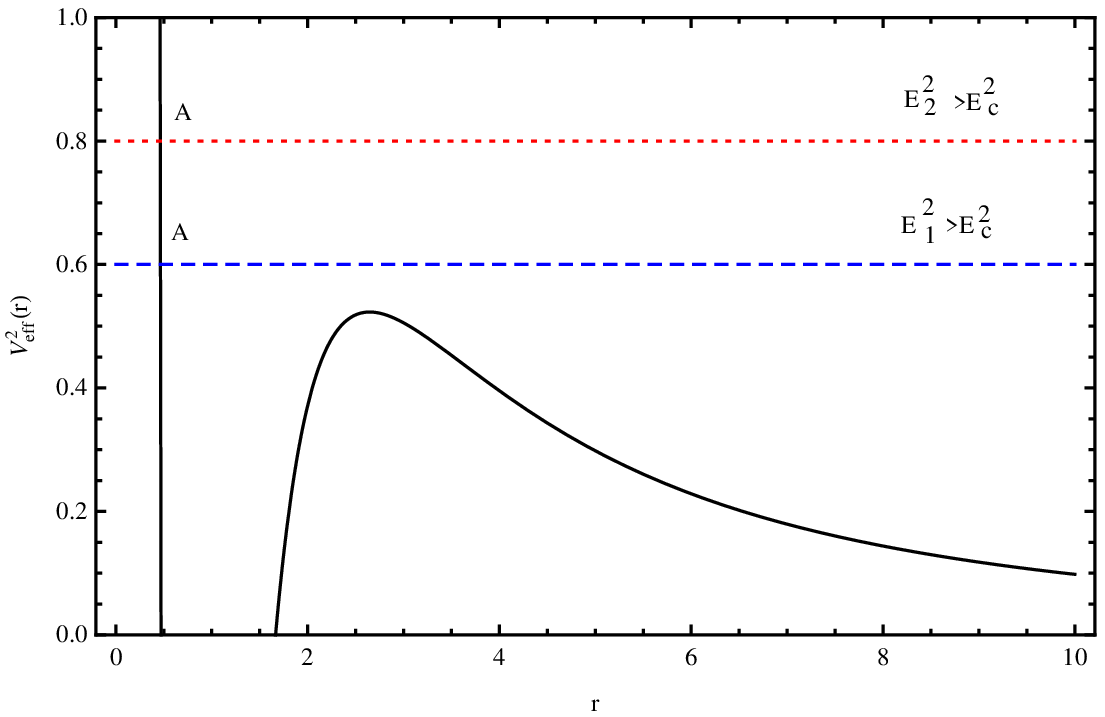}
\includegraphics[scale=0.40]{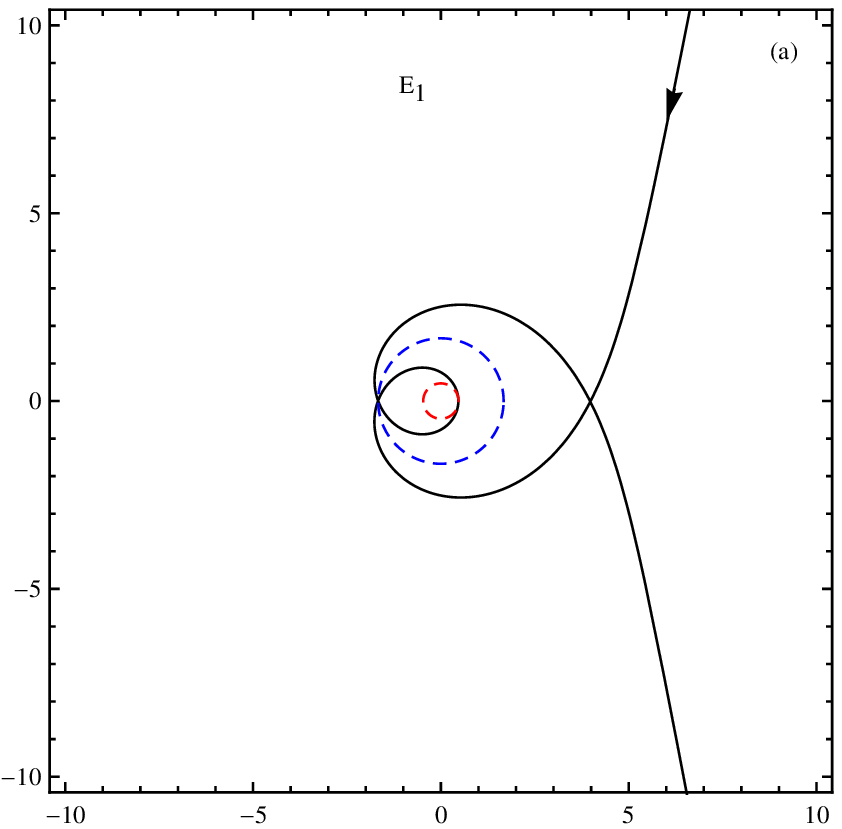}
\includegraphics[scale=0.40]{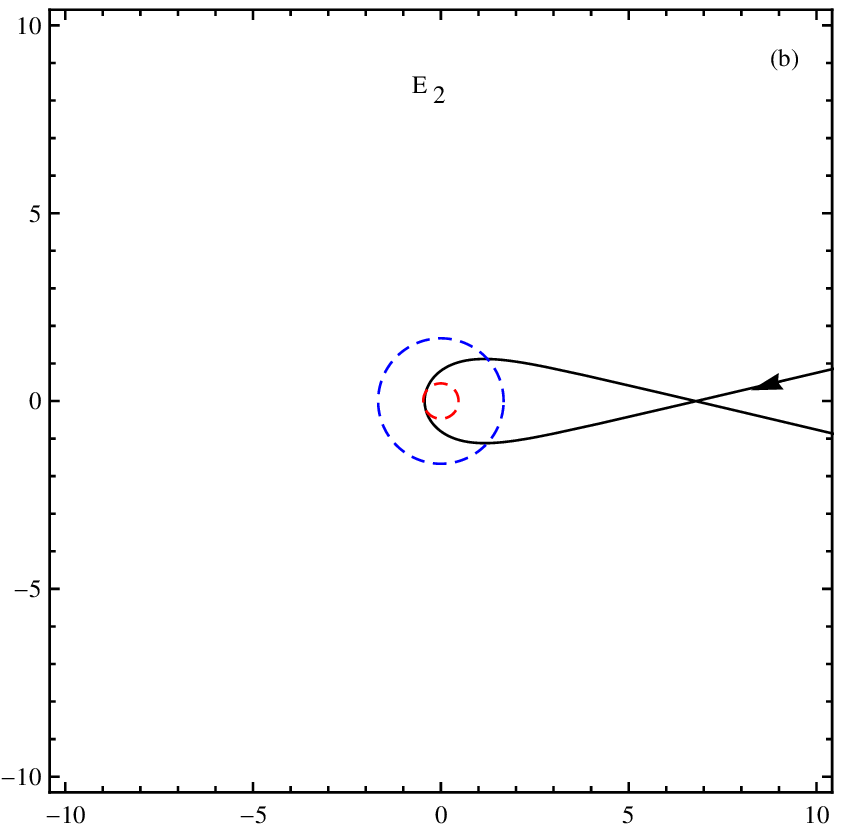}
\includegraphics[scale=0.40]{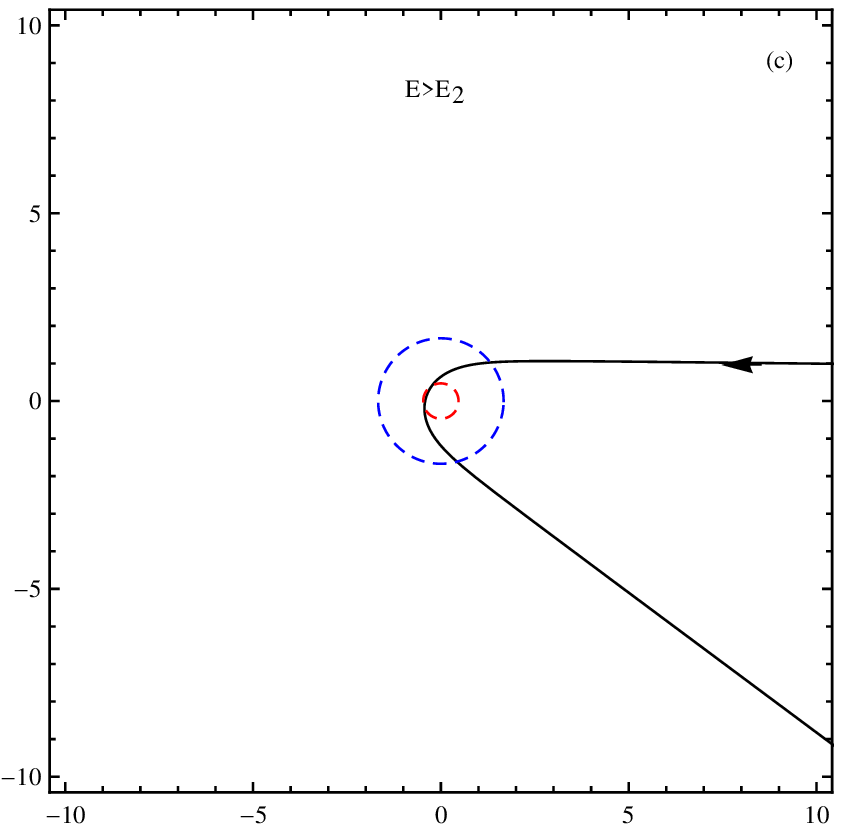}
\caption{Examples of  three kinds of null escape geodesics in  Bardeen spacetime with $g=0.6m$, $L=3.5m$, $m=1$ and two energy levels $E_1^2=0.6$ and $E_2^2=0.8$. }
\end{center}
\end{figure*}

\section{conclusions}
By analyzing the effective potential  of massless and massive test
particles in the Bardeen space-time, which describes a
regular space-time and also represents a singularity free black hole
for $g^{2}<(16/27)m^{2}$, where the parameter $g$ represents the magnetic
charge of the nonlinear self-gravitating monopole, and  numerically simulating  all possible orbits corresponding
to all kinds of  energy levels, we have found that there exist two kinds of
bound orbits, one is close to the center of the black hole and
crosses the two horizons, the other is outside the exterior horizon.
The interesting result is that the planetary orbital precession direction is opposite and heir precession velocities are different, the inner bound orbit
shifts along counter-clockwise with higher velocity while the exterior bound orbit shifts along  clockwise with low velocity,
as shown in Fig.5. We have also found two kinds of circular orbits, the inside one
which closes to the exterior horizon is unstable and the outside one
is a stable circular orbits, and two kinds of escape orbits. For the photon particle, there only exist one many-world bound orbit
which can cross the inner- and out-horizons, one unstable circular orbit and three kinds of escape orbits.

\section{Acknowledgments}
This project is supported by the National Natural Science Foundation
of China under Grant No.10873004, the State Key Development Program
for Basic Research Program of China under Grant No.2010CB832803 and
the Program for Changjiang Scholars and Innovative Research Team in
University, No. IRT0964.


\end{document}